\documentclass[nofootinbib,aps,12pt]{revtex4-1}

\begin{document}

\title{Higgs-Flavor Groups, Naturalness, and Dark Matter}
\author{{\bf S.M. Barr} \\
Bartol Research Institute \\ University of Delaware \\
Newark, Delaware 19716}

\begin{abstract}
In the absence of low-energy supersymmetry, a multiplicity of
weak-scale Higgs doublets would require additional fine-tunings
unless they formed an irreducible multiplet of a non-abelian
symmetry. Remnants of such symmetry typically render some Higgs
fields stable, giving several dark matter particles of various
masses. The non-abelian symmetry also typically gives simple,
testable mass relations.
\end{abstract} \maketitle

The idea that there might be more than one Higgs doublet has been
much investigated for a wide variety of reasons. These reasons
include incorporating CP violation in the scalar sector (as in the
old Weinberg model \cite{weinbergmodel}), addressing the flavor
problem of the quarks and leptons \cite{barr2010, barradeel}), and
obtaining a dark matter candidate, as in ``inert Higgs" models
\cite{inert}.

If (as will be assumed here) there is no low-energy supersymmetry,
then a multiplicity of light Higgs-doublets would exacerbate the
gauge hierarchy problem. Instead of the mass of just one Higgs
doublet being ``fine-tuned" to be much lighter than the Planck
scale, the mass of each Higgs doublet would have to be separately
tuned. The small mass of the Standard Model Higgs doublet relative
to the Planck scale might be accounted for anthropically \cite{abds,
barrkhan, weinberg-scan}, but that does not appear to be the case
for ``extra" Higgs doublets, which do not contribute to the breaking
of the electroweak gauge symmetry. (Throughout this paper,
weak-doublet scalar fields will be called ``Higgs doublets" even if
their vacuum expectation values are zero.)

On the other hand, just one fine-tuning would be sufficient to make
all the Higgs doublets light if all of them were in an irreducible
multiplet of a non-abelian symmetry, $G_{\Phi}$. Then the tuning of
the mass parameter $\mu^2$ of the Standard Model Higgs field would
simultaneously ensure the lightness of all the ``extra" Higgs
doublets, as long as the splitting of the $G_{\Phi}$ multiplet were
small. The breaking of $G_{\Phi}$ can be dynamical, and so this
splitting can be of any magnitude without creating a naturalness
problem.

In short, a multiplicity of elementary Higgs doublets with masses
near the weak scale would seem to require (in the absence of
low-energy supersymmetry) the existence of a non-abelian symmetry
relating their masses. Such a symmetry can have several interesting
consequences, as will be seen in a simple example. It can make some
of the extra Higgs fields stable, giving dark matter candidates, and
give testable relations among the masses of new particles.

If there is such a non-abelian symmetry of the Higgs fields, there
are two possibilities. Either the quarks and leptons of the Standard
Model also transform non-trivially under $G_{\Phi}$, or they are
singlets under $G_{\Phi}$. The former possibility, studied for
example in \cite{barr2010, barradeel} is interesting as an approach
to understanding the flavor structure of the quarks and leptons. It
would generally lead to flavor-changing effects from the exchange of
extra Higgs doublets. Here, however, we study the case where the
only Standard Model field that transforms non-trivially under
$G_{\Phi}$ is the Higgs field. Hence we call such models
``Higgs-flavor models" and the group $G_{\Phi}$ a ``Higgs-flavor
group".

We will show that in typical Higgs-flavor models there exist the
following: (a) One or more extra Higgs doublets that couple to
quarks and leptons proportionally to the Standard Model Higgs
doublet (ensuring ``natural flavor conservation" \cite{nfc}). These
may be light enough to be produced at accelerators. (b) Several
Higgs doublets that do {\it not} couple to quarks and leptons and
that are absolutely stable due to an unbroken discrete subgroup of
the Higgs-flavor symmetry. These are realistic dark matter
candidates, some of which can have masses near present accelerator
limits. (c) Numerous testable symmetry relations among the masses of
the extra Higgs fields. These are of two types: relations among the
masses of different Higgs doublets, and relations among the
$SU(2)_L$-breaking mass splittings within the Higgs doublets.

This phenomenology will be illustrated with a very simple model
based on a Higgs-flavor group $G_{\Phi} = SO(3)$.

In this model, the low-energy theory consists of the Standard Model,
except that the Higgs doublet is part of a 5-plet of $SO(3)_{\Phi}$,
denoted $\Phi^{(ab)}$, where $a$, $b$ are vector indices of
$SO(3)_{\Phi}$ and run from 1 to 3. (The 5-plet is, of course, the
rank-2 symmetric traceless tensor of $SO(3)_{\Phi}$.) There is also
a ``messenger field" $\eta^{(ab)}$, which communicates the effects
of $SO(3)_{\Phi}$ breaking to the Standard Model fields.
$\eta^{(ab)}$ is a real 5-plet of $SO(3)_{\Phi}$, but a singlet
under the Standard Model gauge group. The messenger field is
superheavy, but has a vacuum expectation value (VEV) that is of
order 100 GeV to 1 TeV. This small VEV can arise very simply and in
a technically natural way from the coupling of the messenger field
to the fermions of a sector in which $SO(3)_{\Phi}$ is dynamically
broken by a fermion condensate. (Schematically, if the fermions of
that sector are called $\Psi$ and transform both under
$SO(3)_{\Phi}$ and under an asymptotically free group with a
confinement scale $\Lambda$, one can have terms of the form
$\frac{1}{2} M_{\eta}^2 \eta^2 + f \langle \overline{\Psi} \Psi
\rangle \eta$, which gives $\langle \eta \rangle \sim f
\Lambda^3/M_{\eta}^2$. Even for $M^2_{\eta}$ superheavy, this can be
naturally small.)

Given that the Higgs doublets transform as a 5-plet under the
Higgs-flavor group, whereas the Standard Model quarks and leptons
are singlets under it, the quark and lepton masses must come from
higher-dimension operators, the smallest of which have the form

\begin{equation}
{\cal L}_{yukawa} = Y_{ij} \overline{\psi}_i \psi_j (\Phi^{(ab)}
\eta^{(ab)})/M,
\end{equation}

\noindent where $i$, $j$ are fermion family indices, and repeated
indices of all types are summed over. Such operators can arise from
integrating out vector-like quarks and leptons that carry
$SO(3)_{\Phi}$ indices and have mass of order $\langle \eta
\rangle$, as will be discussed later.

An important point about the term in Eq. (1) is that any Higgs
fields in $\Phi_{(ab)}$ that couple to Standard Model quarks and
leptons do so with the same Yukawa coupling $Y_{ij}$. This
guarantees that the only effects that violate quark and lepton
flavor are through the CKM angles, i.e. ``natural flavor
conservation" (NFC) \cite{nfc}. The reason for NFC in this model is
that the messenger sector is very simple. This is a point to which
weshall return at the end of the paper.

The splittings of the 5-plet of Higgs doublets is given by the
coupling of $\Phi^{(ab)}$ to the messenger field $\eta^{(ab)}$, the
most general renormalizable form of which is given by

\begin{equation}
\begin{array}{rl}
V_2(\Phi^{(ab)}) = & \frac{1}{2} M_{\Phi}^2 \; \Phi^{(ab) \dag}
\cdot \Phi^{(ab)} + \sigma_1 \; \Phi^{(ab) \dag} \cdot \Phi^{(ab)}
\eta^{(cd)} \eta^{(cd)} \\ & \\ & + \sigma_2 \; \Phi^{(ab) \dag}
\cdot \Phi^{(bc)} \eta^{(cd)} \eta^{(da)} +  \sigma_3 \; \Phi^{(ab)
\dag} \cdot \Phi^{(cd)} \eta^{(bc)} \eta^{(da)} \\ & \\ & + \sigma_4
\; \Phi^{(ab) \dag} \cdot \Phi^{(bc)} \eta^{(ab)} \eta^{(cd)} + m_5
\; \Phi^{(ab) \dag} \cdot \Phi^{(bc)} \eta^{(ca)}
\\ & \\
= & \frac{1}{2} M^2 Tr[\Phi^{\dag}_{\lambda} \Phi^{\lambda}] +
\sigma_1 Tr[\Phi^{\dag}_{\lambda} \Phi^{\lambda}]\; Tr[ \eta \eta]
\\ & \\
& + \sigma_2 Tr[\Phi^{\dag}_{\lambda} \Phi^{\lambda} \eta \eta] +
\sigma_3 Tr[\Phi^{\dag}_{\lambda} \eta \Phi^{\lambda} \eta] \\ & \\
& + \sigma_4 Tr[\Phi^{\dag}_{\lambda} \eta] \; Tr[ \Phi^{\lambda}
\eta] + m_5 Tr[\Phi^{\dag}_{\lambda} \Phi^{\lambda} \eta],
\end{array}
\end{equation}

\noindent where the coefficients are real. In the first expression
for $V_2$, the dot represents the contraction of $SU(2)_L$ indices,
which are not shown. In the second expression, the $SU(2)_L$ indices
are denoted by $\lambda$, and the traces are over the $SO(3)_{\Phi}$
indices, which are not shown. The terms in Eq. (2) are actually not
all independent. In terms of the form $(\Phi^{\dag} \Phi)(\eta
\eta)$, the product of $\eta$ with itself must be in the symmetric
product $(5 \times 5)_S = 1 + 5 + 9$. Thus the four terms with
coefficients $\sigma_i$ in Eq. (2) depend on only three invariant
combinations.

Because $\langle \eta^{(ab)} \rangle$ is a real symmetric matrix, an
$SO(3)_{\Phi}$ basis can be chosen where it is real and diagonal.
Without loss of generality, then, the VEV of the messenger field may
be written

\begin{equation}
\langle \eta^{(ab)} \rangle = \left( \begin{array}{ccc} a +
\frac{1}{\sqrt{3}} b & 0 & 0 \\
0 & - a + \frac{1}{\sqrt{3}} b & 0 \\
0 & 0 & - \frac{2}{\sqrt{3}} b \end{array} \right).
\end{equation}

\noindent Let us write the components of the 5-plet of Higgs
doublets as

\begin{equation}
\Phi^{(ab)} = \left( \begin{array}{ccc} A +
\frac{1}{\sqrt{3}} B & \Phi^{(12)} & \Phi^{(13)} \\
\Phi^{(12)} & - A + \frac{1}{\sqrt{3}} B & \Phi^{(23)} \\
\Phi^{(13)} & \Phi^{(23)} & - \frac{2}{\sqrt{3}} B \end{array}
\right).
\end{equation}

\noindent Then it is directly seen from Eq. (1) that the doublet
that couples to the known quarks and leptons is the linear
combination $\Phi^{(ab)} \langle \eta^{(ab)} \rangle = 2 \sqrt{a^2 +
b^2} \Phi_+$, where $\Phi_+ \equiv \frac{1}{\sqrt{a^2 + b^2}} (aA +
bB)$. The orthogonal combination will be denoted $\Phi_- \equiv
\frac{1}{\sqrt{a^2 + b^2}} (bA - aB)$. We shall call $\Phi_+$ and
$\Phi_-$ the ``diagonal Higgs doublets", and $\Phi^{(ab)}$ with $a
\neq b$ the ``off-diagonal Higgs doublets". (Note that we define
these with respect to the basis in which the VEV of the messenger
field has the form given in Eq. (3).)

The mass spectrum of the doublets that results from Eq. (2) is
easily computed by substituting into it the forms given in Eqs. (3)
and (4). Let us first write it in the simple case where the cubic
term in Eq. (2) vanishes exactly, i.e. $m_5 = 0$. (This could arise
from a $Z_2$ symmetry under which the messenger fields are odd.)
Then we will consider the slightly more interesting case where $m_5
\neq 0$. If $m_5 =0$, then the mass eigenstates and eigenvalues come
out to be the following:

\begin{equation}
\begin{array}{cl}
M^2(\Phi^{(12)}) & = M_0^2 + (\sigma_2 - \sigma_3) \; 2 a^2 +
(\sigma_2 + \sigma_3) \; \frac{2}{3} b^2, \\
&
\\ M^2(\Phi^{(13)}) & = M_0^2 + \sigma_2 \; a^2 + (\frac{5}{3} \sigma_2 -
\frac{4}{3} \sigma_3) \; b^2 + (\sigma_2 - 2 \sigma_3) \; \frac{2}{\sqrt{3}} ab, \\
&
\\ M^2(\Phi^{(23)}) & = M_0^2 + \sigma_2 \; a^2 + (\frac{5}{3} \sigma_2 -
\frac{4}{3} \sigma_3) \; b^2 - (\sigma_2 - 2 \sigma_3) \;
\frac{2}{\sqrt{3}} ab,
\\ &
\\ M^2(\Phi_-) & = M_0^2 + (\sigma_2 + \sigma_3) \; \frac{2}{3} (a^2 +
b^2), \\ & \\ M^2(\Phi_+) & = M_0^2 + (\sigma_2 + \sigma_3 + 2
\sigma_4)\; 2(a^2 + b^2),
\end{array}
\end{equation}

\noindent where $M_0^2 \equiv M_{\Phi}^2 + 4 \sigma_1 (a^2 + b^2)$.
(We are at the moment neglecting $SU(2)_L$-breaking effects.) One
sees that the combination that couples to quarks and leptons,
$\Phi_+$, is actually a mass eigenstate. If the quarks and leptons
are to obtain mass, then $\Phi_+$ must obtain a VEV and must
therefore be identified with the Standard Model Higgs doublet (which
we will call $\Phi_{SM}$) and have a negative mass-squared with
magnitude of order (100 GeV)$^2$. The other four Higgs doublets must
have positive mass-squared large enough to evade present limits.
Therefore, $\Phi_+$ must have the smallest mass-squared in Eq. (5),
which can happen if, for example, $\sigma_4$ is sufficiently
negative. The magnitude and sign of $M^2(\Phi_+)$ has to be
explained anthropically, presumably in the context of a multiverse
scenario. It is simplest to imagine that the parameter that varies
or ``scans" among the domains of the multiverse \cite{weinberg-scan}
is the $SO(3)_{\Phi}$-invariant mass parameter $M_{\Phi}^2$ of the
5-plet.

From Eq. (5), one finds a simple experimentally testable
relationship among the mass splittings of the four extra Higgs
doublets $\Phi_-$, $\Phi^{(12)}$, $\Phi^{(13)}$, and $\Phi^{(23)}$.
One has

\begin{equation}
\begin{array}{cl}
\Delta^2_{12} & \equiv M^2(\Phi^{(23)}) - M^2 (\Phi_-) = \frac{4}{3}
(\sigma_2 - 2 \sigma_3) \; a^2, \\ \Delta^2_{13} & \equiv
M^2(\Phi^{(12)}) - M^2 (\Phi_-) = (\sigma_2 - 2
\sigma_3) \left( \frac{1}{\sqrt{3}} a + b \right)^2, \\
\Delta^2_{23} & \equiv M^2(\Phi^{(13)}) - M^2 (\Phi_-) = (\sigma_2 -
2 \sigma_3) \left( \frac{1}{\sqrt{3}} a - b \right)^2, \end{array}
\end{equation}

\noindent from which one sees that

\begin{equation}
\begin{array}{l}
|\Delta^2_{12}|^{1/2} = |\Delta^2_{13}|^{1/2} +
|\Delta^2_{23}|^{1/2}, \;\; {\rm if} \;\; |b| < \frac{1}{\sqrt{3}}
|a|, \\
|\Delta^2_{13}|^{1/2} = |\Delta^2_{12}|^{1/2} +
|\Delta^2_{23}|^{1/2}, \;\; {\rm if} \;\; |b| > \frac{1}{\sqrt{3}}
|a|, \;\; ab >0, \\
|\Delta^2_{23}|^{1/2} = |\Delta^2_{12}|^{1/2} +
|\Delta^2_{13}|^{1/2}, \;\; {\rm if} \;\; |b| > \frac{1}{\sqrt{3}}
|a|, \;\; ab <0.
\end{array}
\end{equation}

\noindent In other words, the largest of the three mass-squared
splittings among the extra Higgs doublets is simply related to the
other two. Another relation implied by Eq. (6) is that $\Phi_-$ is
either heavier than all the off-diagonal Higgs doublets or lighter
than them all, depending on the sign of $\sigma_2 - 2 \sigma_3$.

Eqs. (6) and (7) allow us to draw some conclusions about the
stability of the four extra Higgs doublets. In what follows, when we
say that a doublet is stable, we mean that its lightest component is
stable, since the heavier components within a weak doublet can decay
into lighter ones by charged weak interactions.

First, consider the lightest two of the three off-diagonal Higgs
doublets. These are rendered absolutely stable by unbroken discrete
subgroups of the Higgs-flavor symmetry. (And this conclusion applies
even if $m_5 \neq 0$.) The relevant symmetries are the parity
transformations $P_a$, where $P_a$ reverses the sign of the $a^{th}$
component of an $SO(3)_{\Phi}$ vector. With respect to $P_a$, any
field with an odd(even) number of $SO(3)_{\Phi}$ indices equal to
$a$ is odd(even). For example, $\Phi^{(12)}$ and $\Phi^{(13)}$ are
odd under $P_1$, while $\Phi^{(23)}$ is even. These parities are
unbroken by the VEV in Eq. (3). Thus the lightest $P_a$-odd fields
are stable. For example, if $\Phi^{(12)}$ is the heaviest of the
off-diagonal Higgs doublets, then $\Phi^{(13)}$ is the lightest
$P_1$-odd multiplet and $\Phi^{(23)}$ is the lightest $P_2$-odd
multiplet. Therefore the lightest components of $\Phi^{(13)}$ and
$\Phi^{(23)}$ are absolutely stable. These will contribute to the
dark energy of the universe as will be discussed briefly later.

The heaviest of the three off-diagonal Higgs doublets is not
prevented by these discrete symmetries from decaying into lighter
Higgs doublets. Indeed quartic terms exist which would appear to
allow such decays (for example, $\Phi^{(ab) \dag} \cdot \Phi^{(bc)}
\Phi^{(cd) \dag} \cdot \Phi^{(da)}$, which contains $\Phi^{(12)
\dag} \cdot \Phi^{(23)} \Phi^{(31) \dag} \cdot \Phi_{\pm}$). Whether
such decays can occur depends on kinematics. One must consider
separately two cases. As noted above, $\Phi_-$ is either the
lightest or the heaviest of the four extra Higgs doublets, depending
on the sign of $(\sigma_2 - 2 \sigma_3)$. Call these Cases I and II
respectively.

{\bf Case I}. If $\Phi_-$ is the lightest of the extra Higgs
doublets, the decay of the heaviest off-diagonal Higgs doublet into
lighter Higgs doublets is kinematically forbidden, as we will now
show. (We are still neglecting the $SU(2)_L$-breaking contributions
to the masses of the extra Higgs doublets.) If the mass-squared of
$\Phi_-$, which is positive, is denoted $m_0^2$, then by Eq. (7) the
mass-squareds of the three off-diagonal Higgs doublets can be
written (in ascending order) as $m_0^2 + x^2$, $m_0^2 + y^2$, and
$m_0^2 + (x + y)^2$ for some $x$ and $y$ such that $y > x > 0$ . For
the heaviest off-diagonal Higgs doublet to decay into the two
lightest off-diagonal Higgs doublets plus other particles (which is
the only decay allowed it by the parity symmetries $P_a$), one must
have $\sqrt{m_0^2 + (x+y)^2}
> \sqrt{m_0^2 + x^2} + \sqrt{m_0^2 + y^2} > \sqrt{m_0^2 + x^2} + y$
$\Rightarrow m_0^2 + (x+y)^2 > m_0^2 + x^2 + y^2 + 2 y \sqrt{m_0^2 +
x^2}$ $\Rightarrow x > \sqrt{m_0^2 + x^2}$,  which is clearly
impossible since $m_0^2
>0$ by the fact that all the extra Higgs doublets have positive
mass-squared. The heaviest off-diagonal Higgs doublet also cannot
decay directly into quarks and leptons, since it has no Yukawa
coupling to them. It is therefore stable.

In Case I, the Higgs doublet $\Phi_-$ is also stable, because it has
no Yukawa couplings to the quarks and leptons, and because there
turn out to be no quartic couplings that allow its decay into three
$\Phi_+$. Some of these conclusions are modified if $m_5 \neq 0$ as
will be seen.

{\bf Case II}. In Case II, whether $\Phi_-$ and the heaviest
off-diagonal Higgs doublet are able to decay into lighter Higgs
doublets depends on the values of parameters. The lightest two
off-diagonal Higgs doublets are, however, absolutely stable due to
the symmetries $P_a$ (that is, their lightest components are).

This model differs from most models of dark matter in that there are
several stable particles that contribute to the dark matter density
of the universe. Most of the dark matter density would come from the
heaviest stable extra Higgs particle, because of both its smaller
annihilation cross-section and the larger mass. However, in the
present model, the lightest stable extra Higgs particle can be much
lighter than heaviest one. For example, in Case I, $\Phi_-$ can be
much lighter than the heaviest stable Higgs doublet. In Case II, the
lightest off-diagonal Higgs doublet can be much lighter than all
three of the other extra Higgs doublets. Thus, even if the particle
which is the dominant component of the dark matter has a mass of
order a TeV, there can be other stable Higgs particles several times
lighter than that. The calculation of the dark matter density is
obviously quite involved as it depends on the eleven parameters of
Eqs. (2) and (12), and will be considered in detail elsewhere.

Now let us consider the model when $m_5 \neq 0$. This term in Eq.
(2) has no affect on the masses of the off-diagonal Higgs doublets,
but modifies the $2 \times 2$ mass matrix of the diagonal Higgs
doublets, so that the eigenstates are mixtures of what we called
$\Phi_{\pm}$:

\begin{equation}
\begin{array}{ccl}
\Phi_{SM} = & \Phi'_+ & = \cos \theta_H \Phi_+ - \sin \theta_H \Phi_-, \\
& & \\
& \Phi'_- & = \sin \theta_H \Phi_+ + \cos \theta_H \Phi_-,
\end{array}
\end{equation}

\noindent where, for small $m_5$,

\begin{equation}
\tan \theta_H \cong \frac{m_5 a (-a^2 + 3 b^2)/(a^2 + b^2)}
{M^2(\Phi'_-) - M^2(\Phi'_+)}.
\end{equation}

\noindent Thus the diagonal extra Higgs doublet $\Phi'_-$ now
couples to the quarks and leptons with a strength that is simply
$\tan \theta_H$ times that of the Standard Model Higgs doublet and
is no longer stable. From the decays of $\Phi'_-$ into quarks, the
value $\tan \theta$ is in principle directly measurable.

There is also a shift in the mass of $\Phi'_-$ from the value
predicted in Eq. (7). For small $m_5$, this shift is given by

\begin{equation}
\delta M^2(\Phi_-) = - \frac{2}{\sqrt{3}} m_5 b \frac{3a^2 -
b^2}{a^2 + b^2}.
\end{equation}

\noindent Thus one has the prediction

\begin{equation}
\delta M^2(\Phi_-) = - r \frac{3 - r^2}{3r^2-1} \tan \theta_H \left[
M^2(\Phi_-) - M^2(\Phi_+) \right],
\end{equation}

\noindent where $r \equiv b/a$ can be extracted from Eqs. (6) and
(7). In particular $\left| (r + \frac{1}{\sqrt{3}})/(r -
\frac{1}{\sqrt{3}}) \right| = \sqrt{\Delta^2_{13}/\Delta^2_{23}}$.
This shifting of the mass of $\Phi_-$ has the effect that in Case I
for certain values of parameters the heaviest off-diagonal Higgs
boson can decay into other Higgs doublets.

Up to this point, we have neglected the $SU(2)_L$-breaking effects
in computing the masses of the extra Higgs doublets. This breaking
gives splitting with each doublet between the charged, neutral
scalar, and neutral pseudoscalar components. The splitting occurs
through the coupling of the Standard Model Higgs doublet to the
extra Higgs doublets in the quartic terms in the Higgs potential.
Since the Higgs-flavor group $SO(3)_{\Phi}$ significantly constrains
the form of the those quartic terms, testable predictions arise for
the pattern of $SU(2)_L$-breaking splittings. The most general form
for the quartic part of the Higgs potential is

\begin{equation}
\begin{array}{rl}
V_4 (\Phi^{(ab)}) = &  \lambda_1 \; \left[ \Phi^{(ab) \dag} \cdot
\Phi^{(ab)} \right]^2 + \lambda_2 \; \left[ \Phi^{(ab) \dag} \cdot
\Phi^{(cd)} \right] \left[ \Phi^{(cd) \dag} \cdot \Phi^{(ab)}
\right] \\ & + \lambda_3 \; \left[ \Phi^{(ab) \dag} \cdot
\Phi^{(cd)} \right] \left[ \Phi^{(ab) \dag} \cdot \Phi^{(cd)}
\right] + \lambda_4 \; \left[ \Phi^{(ab) \dag} \cdot \Phi^{(cd)}
\right] \left[ \Phi^{(ac) \dag} \cdot \Phi^{(bd)} \right] \\ & +
\lambda_5 \; \left[ \Phi^{(ab) \dag} \cdot \Phi^{(bc)} \right]
\left[ \Phi^{(cd) \dag} \cdot \Phi^{(da)} \right] + \lambda_6 \;
\left[ \Phi^{(ab) \dag} \cdot \Phi^{(bc)} \right] \left[ \Phi^{(da)
\dag} \cdot \Phi^{(cd)} \right] \\ & \\
= & \lambda_1 \left[ Tr \left( \Phi^{\dag}_{\lambda} \Phi^{\lambda}
\right) \right]^2 + \lambda_2 \; Tr \left[ \Phi^{\dag}_{\lambda}
\Phi^{\mu} \right] Tr \left[ \Phi^{\dag}_{\mu} \Phi^{\lambda}
\right] \\
& + \lambda_3 \; Tr \left[ \Phi^{\dag}_{\lambda} \Phi^{\dag}_{\mu}
\right] Tr \left[ \Phi^{\lambda} \Phi^{\mu} \right] + \lambda_4 \;
Tr \left[ \Phi^{\dag}_{\lambda} \Phi^{\mu} \Phi^{\lambda}
\Phi^{\dag}_{\mu} \right] \\
& + \lambda_5 \; Tr \left[ \Phi^{\dag}_{\lambda} \Phi^{\lambda}
\Phi^{\dag}_{\mu} \Phi^{\mu} \right] + \lambda_6 \; Tr \left[
\Phi^{\dag}_{\lambda} \Phi^{\lambda} \Phi^{\mu} \Phi^{\dag}_{\mu}
\right]
\end{array}
\end{equation}

\noindent where the notation is the same as in Eq. (2). By the
hermiticity of $V_4$ the $\lambda_i$ are real. Since the product
$(\Phi^{\dag} \Phi)$ must be in $5 \times 5 = 1 + 3 + 5 + 7 + 9$,
the six terms in Eq. (12) depend on only five invariant
combinations.

It is easy to show that $V_4$ given in Eq. (12) contains no terms of
the form $\Phi^{\dag}_- \Phi_+ \Phi^{\dag}_+ \Phi_+$. Consequently,
when $\Phi_+$ acquires a VEV it does not mix $\Phi_-$ and $\Phi_+$.
Therefore, the conclusion reached earlier that for $m_5 = 0$ the
doublet $\Phi_-$ does not couple to quarks and leptons still holds.

When $\Phi_{SM}$ acquires a vacuum expectation value, the terms in
Eq. (12) (except for the $\lambda_1$ term) give $SU(2)_L$-breaking
contributions to the masses of the Higgs fields. Each Higgs doublet
has a charged, neutral scalar, and neutral pseudoscalar component,
and thus two splittings. So the four extra Higgs doublets have
altogether eight $SU(2)_L$-breaking splittings, which are determined
by the five parameters $\lambda_i$, $i = 2, ..., 6$ (of which only
four are independent of each other). There are therefore four
testable mass relations.

There are some technical points about symmetry breaking to be
considered. We analyzed two cases $m_5 =0$ and $m_5 \neq 0$. The
case $m_5 =0 $ can be realized if there is a $Z_2$ under which the
messenger field $\eta^{(ab)}$ is odd and all the Standard Model
fields are even. The sector that dynamically breaks $SO(3)_{\Phi}$
could then (for example) have the form $\frac{1}{2} M^2_{\eta}
\eta^{(ab)} \eta^{(ab)} + \sum_{I = 1}^5 f_I (\overline{\Psi}^{(ab)}
\Psi_I) \eta^{(ab)}$, where under $SU(N)_{DSB} \times SO(3)_{\Phi}
\times Z_2$, one has $\overline{\Psi}^{(ab)} = (\overline{N}, 5,
+)$, $\Psi_I = (N, 1, -)$, and $\eta^{(ab)} = (1, 5, -)$. The
confining group $SU(N)_{DSB}$ causes a $\overline{\Psi} \Psi$
condensate to form that induces a linear term, and thus a VEV, for
the messenger field. In the case $m_5 \neq 0$ one must explain why
$m_5$ is of order the weak scale rather than the Planck scale. Here
too one can invoke a $Z_2$. In this case, one could have two
messenger fields, $\eta^{(ab)}$ and $\eta'$ that are respectively a
5-plet and a singlet under $SO(3)_{\Phi}$ and that are both odd
under $Z_2$. The dynamical symmetry breaking sector could (for
example) have the form $\frac{1}{2} M^2_{\eta} \eta^{(ab)}
\eta^{(ab)} + \frac{1}{2} M^2_{\eta'} \eta' \eta' + \sum_{I = 1}^5
f_I (\overline{\Psi}^{(ab)} \Psi_I) \eta^{(ab)} + f'
(\overline{\Psi}' \Psi') \eta'$. If the confining scale of
$SU(N)_{DSB}$ is $\Lambda$, and both $M^2_{\eta} \sim M^2_{\eta'}$
are of superheavy scale, then both the 5-plet and singlet messenger
fields will naturally have VEVs of the same order of magnitude,
namely $\Lambda^3/M^2_{\eta}$.

The effective Yukawa operators given in Eq. (1) can arise through
integrating out fermions that have mass of order $\langle \eta
\rangle$.  For example, if there is a set of fermions $\psi^{(ab)} =
(1, 5, +)$ that has the quantum numbers of a family under the
Standard Model gauge group, and $\overline{\psi}^{(ab)} = (1, 5, -)$
that has the quantum numbers of an anti-family, then there can be
renormalizable Yukawa couplings of the form $\psi_i \psi^{(ab)}
\Phi^{(ab)}$, $\overline{\psi}^{(ab)} \psi^{(ab)} \eta'$, and
$\overline{\psi}^{(ab)} \psi_j \eta^{(ab)}$. When the $\psi^{(ab)} +
\overline{\psi}^{(ab)}$ is integrated out, it leads at tree level to
effective terms of the form $\psi_i \psi_j (\Phi^{(ab)}
\eta^{(ab)}/\langle \eta' \rangle$, as given in Eq. (1). As they
arise at tree level, there is no reason why some of the coefficients
of such effective terms could not be of order one (as would be
needed for the $t$ quark mass).

One further point: the Higgs-flavor group can be local. The gauge
bosons of $G_{\Phi}$ would obtain mass of order the condensate
$\Lambda \sim (\langle \eta \rangle M^2_{\eta})^{1/3} \gg \langle
\eta \rangle$.  Therefore, they would have negligible effect at low
energies.

We conclude by noting that the ``Higgs flavor" model presented here
is typical but hardly unique. There are different possibilities for
the non-abelian Higgs-flavor group $G_{\Phi}$ (including both
continuous and discrete groups), and various possibilities for the
$G_{\Phi}$ representations for the Higgs doublets and messenger
fields. One would expect, however, that typical features of such
models would include the existence of one or more absolutely stable
extra Higgs doublets that contribute to dark matter and some of
which can be quite light, and the existence of other Higgs doublets
that couple to the known quarks and leptons proportionally to the
Standard Model Higgs.

One expects these features to be typical because they tend to follow
from having a very simple messenger sector, and a simple messenger
sector is required to ensure ``natural flavor conservation" of the
quarks and leptons. For example, in the model described in this
paper, if there were two messenger fields, $\eta^{(ab)}$ and
$\eta^{\prime (ab)}$, instead of just one, then instead of the
single Yukawa term of Eq. (1), there would be two: ${\cal
L}_{yukawa} = Y_{ij} \overline{\psi}_i \psi_j (\Phi^{(ab)}
\eta^{(ab)})/M + Y'_{ij} \overline{\psi}_i \psi_j (\Phi^{(ab)}
\eta^{\prime (ab)})/M$. Since $Y_{ij}$ and $Y'_{ij}$ would have no
reason to be simultaneously diagonalizable, potentially large quark
and lepton flavor-changing mediated by scalar exchange would result.
Ensuring natural flavor conservation in Higgs-flavor models thus
generally requires a minimal set of messenger fields. But this in
turn makes the $G_{\Phi}$-breaking in the low-energy Higgs sector
very simple and tends to leave unbroken in that sector a discrete
remnant of the Higgs-flavor symmetry that can render some of the
Higgs fields absolutely stable, as we have seen.

The general lesson, then, is that the non-abelian Higgs-flavor
symmetry required to make extra elementary Higgs doublets naturally
light in the absence of low-energy supersymmetry, together with the
simplicity of the messenger sector needed to avoid excessive quark
and lepton flavor changing, tends to result in stable Higgs
particles that can play the role of dark matter. It also can give
rise to unstable extra Higgs fields that couple to quarks and
leptons proportionally to the Standard Model Higgs field. And,
finally, it tends to yield simple and testable mass relations among
the extra Higgs fields.


\begin{thebibliography}{999}
\bibitem{weinbergmodel} S. Weinberg, {\it Phys. Rev. Lett.} {\bf
37}, 657 (1976).
\bibitem{barr2010} S.M. Barr, {\it Phys. Rev.} {\bf D82}, 055010
(2010).
\bibitem{barradeel} M. A. Ajaib and S.M. Barr, hep/...
\bibitem{inert} E. Ma, {\it Phys. Rev.} {\bf D73}, 077301 (2006);
R. Barbieri, L.J. Hall, and V.S. Rychkov, {\it Phys. Rev.} {\bf
D74}, 015007 (2006); M. Cirelli, N. Fornengo, and A. Strumia, {\it
Nucl. Phys.} {\bf B753}, 178 (2006).
\bibitem{abds} V. Agrawal, S.M. Barr, J.F. Donoghue, and D. Seckel,
{\it Phys. Rev.} {\bf D57}, 5480 (1998); {\it ibid.} {\it Phys. Rev.
Lett.} {\bf 80} 1822 (1998).
\bibitem{barrkhan} S.M. Barr and Almas Khan, {\it Phys. Rev.} {\bf
D76}, 045002 (2007).
\bibitem{weinberg-scan} S. Weinberg, ``Living in the Multiverse", in
{\it Universe or Multiverse?}, ed. B.J. Carr (Cambridge Univ. Press,
Cambridge, 2007).
\bibitem{nfc} S.L. Glashow and S. Weinberg, {\it Phys. Rev.} {\bf
D15}, 1958 (1977).
\end{thebibliography}
\end{document}